\documentclass[aps,pra,reprint,superscriptaddress,showpacs,longbibliography,floatfix,footnoteinbib]{revtex4-1}
\usepackage{amsmath,amssymb,graphicx,units}
\usepackage[plainpages=false,pdfpagelabels,colorlinks=true,linkcolor=red,urlcolor=blue,citecolor=blue,pdftitle={Title},pdfauthor={},pdfdisplaydoctitle=true,pdfduplex=DuplexFlipLongEdge]{hyperref}
\usepackage{multirow}
\usepackage{setspace}

\begin{document}

\title{Enhanced superconductivity and various edge modes in modulated $t$-$J$ chains}

\author{Yong-Feng Yang}
\affiliation{School of Physical Science and Technology $\&$ Lanzhou Center for Theoretical Physics $\&$ Key Laboratory of Theoretical Physics of Gansu Province, Lanzhou University, Lanzhou, Gansu 730000, China}

\author{Jing Chen}
\affiliation{Center for Computational Quantum Physics, Flatiron Institute, New York, NY 10010 USA}

\author{Chen Cheng}
\email{chengchen@lzu.edu.cn}
\affiliation{School of Physical Science and Technology $\&$ Lanzhou Center for Theoretical Physics $\&$ Key Laboratory of Theoretical Physics of Gansu Province, Lanzhou University, Lanzhou, Gansu 730000, China}
\affiliation{Beijing Computational Science Research Center, Beijing 100094, China}

\author{Hong-Gang Luo}
\email{luohg@lzu.edu.cn}
\affiliation{School of Physical Science and Technology $\&$ Lanzhou Center for Theoretical Physics $\&$ Key Laboratory of Theoretical Physics of Gansu Province, Lanzhou University, Lanzhou, Gansu 730000, China}
\affiliation{Beijing Computational Science Research Center, Beijing 100094, China}

\begin{abstract}

We numerically investigate the ground state of the extended $t$-$J$ Hamiltonian with periodic local modulations in one dimension by using the density-matrix renormalization group method. Examining charge and spin excitation gaps, as well as the pair binding energy, with extrapolated results to the thermodynamic limit, we obtain a rich ground-state phase diagram consisting of the metallic state, the superconducting state, the phase separation, and insulating states at commensurate fillings. Compared to the homogeneous 1D $t$-$J$ model, the superconductivity is greatly enhanced and stabilized by the flat-band structure. This superconducting state in quasi-periodic chains shares similar properties with ladder systems: significant negative pair binding energy occurs, and the singlet pairing correlation function dominates with the algebraic decay while the single-particle Green's function and spin correlation function decay exponentially. On the other hand, quasi-periodicity leads to nontrivial topological nature in insulating states, characterized by different integer Chern numbers at different fillings. Due to the interplay among the topology, the interaction, and the 1D confinement, gapless edge modes show strong spin-charge separation and in different regions can relate to different collective modes, which are the charge of a single fermion, the magnon, and the singlet-pair. We also find two interaction driven topological transitions: i) at particle filling $\rho=1/2$, the low-energy edge excitations change from the magnon to singlet-pair, accompanied with pair formation in bulk; and ii) at $\rho=3/4$, while the gapless edge mode remains the charge of a single fermion, there is a gap-closing point and a $\pi$-phase shift in the quasi-particle spectrum. 
\end{abstract}

\maketitle

\section{Introduction}

The $t$-$J$ Hamiltonian~\cite{Zhang1988} is one of the most important theoretical models in strongly correlated systems, especially for its role as a canonical model for studying high-temperature superconductors~\cite{Dagotto1994,Keimer2015}. While the initial interest of the model resides in its two-dimensional (2D) realization, the one-dimensional (1D) $t$-$J$ model attracts many investigations as it shares some signatures with the 2D case like the spin-gap and the superconducting phase~\cite{Ogata2008,Moreno2011}. However, the superconducting state in the 1D chain supports a very weak spin gap and pair binding, which vanishes for physically relevant values of $J$ in real materials~\cite{Moreno2011,Zhu2014}. Compared to the pure 1D case, in a higher dimension, even-leg $t$-$J$ ladders show stronger evidence of superconductivity with significant binding energy~\cite{Dagotto1992,Hayward1995,Hayward1996,Poilblanc2003,White1997,Zhu2014} and have been extensively studied~\cite{Cheng2018,Jiang2021,Gong2021,Jiangst2021}. In a recent work~\cite{Reja2016}, the substantial negative binding energy has been demonstrated in 1D coupled $t$-$J$ segments for physically relevant values of exchange $J$ and hole doping. This work provides new possibilities in superconducting material design and raises whether the enhanced superconductivity is pervasive for general quasi-periodic systems. 

The 1D quasi-periodic superlattice has been attracting continuous investigations in the last decade, partially because of its nontrivial topological nature~\cite{Lang2012,Kraus2012,Xu2013,Hu2016,Zhu2013,Kuno2017,Stenzel2019}. In a noninteracting setting, the essential physics behind it is well understood, that these 1D quasi-periodic models can be mapped to the 2D Harper-Hofstadter model~\cite{Harper1955,Hofstadter1976} which is used to describe the quantum Hall effect~\cite{Klitzing2017}. In the presence of interactions, there occur more unique phenomena such as fractional topological insulators~\cite{Xu2013,Hu2016} and topological Mott insulators~\cite{Zhu2013,Kuno2017,Stenzel2019}. Moreover, in the spinful fermionic Hubbard chain, the spin-charge separation due to the confined geometry can also affect topological properties, leading to the phase transition with the low-energy edge excitations change from spin-1/2 fermionic single-particle modes to spin-1 bosonic collective modes at specific fillings~\cite{Hu2019}. 

The quasi-periodic $t$-$J$ Hamiltonian in 1D superlattices has plentiful elements supporting pregnant physics, which has not been systematically explored. The interplay among topology, interaction, spin-charge separation, and pairing would bring interesting phenomena in perspectives from both bulk and edge properties, such as the enhanced superconductivity and various topological edge modes. In parallel, the development of ultracold gases in optical lattices within the last two decades provides a new way to explore the physical phenomena described by theoretical model Hamiltonians~\cite{BlochRMP2008,esslinger_review_10, Ni231, Park2015, DeMarco2019}, including quasi-periodic models~\cite{Tai2017} and extended $t$-$J$ models~\cite{Gorshkov2011, Gorshkov2011a}. The experimental setting with remarkable control and tunability has also brought more flexibilities to the theoretical model, which can be studied with sufficient motivation in a broader range of parameters and in different extended forms~\cite{Gorshkov2011, Gorshkov2011a, Cheng2015, Fazzini2019}. The quasi-periodic $t$-$J$ Hamiltonian, which has not yet been explicitly proposed in optical lattices, has the potential to be realized in future experiments~\cite{Jepsen2020}. Motivated by various possibilities, in this work, we systematically investigate the $t$-$J$ model in 1D superlattices with quasi-periodic modulations in a wide range of parameters. We aim to draw the ground-state phase diagram of the modulated $t$-$J$ chain and unveil its nontrivial topological nature from numerically exact results. 

The remainder of the paper is organized as follows. In Sec.~\ref{sec:model}, we introduce the quasi-periodic $t$-$J$ model, the numerical method, and the way to characterize the bulk and topological properties. In Sec.~\ref{sec:bulk}, we summarize our main results by the ground-state phase diagram and introduce the bulk state properties, especially the enhanced superconductivity. In Sec.~\ref{sec:edge} we focus on the topological nature of insulating states and demonstrate various gapless edge modes corresponding to different quasi-particles. Finally, the summary and discussion are made in Sec.~\ref{sec:summary}. 

\section{Model and Methodology}
\label{sec:model}

We investigate the extended $t$-$J$ model with site-dependent couplings and interactions,
\begin{align}
{\cal H}=&\sum_{i\sigma}t_{i}\left(\hat{c}_{i,\sigma}^{\dagger}\hat{c}_{i+1,\sigma}+\mathrm{H.c.}\right) \nonumber \\
&+\sum_{i}J_{i}\left(\vec{S}_{i}\cdot\vec{S}_{i+1}-\frac{1}{4}\hat{n}_{i}\hat{n}_{i+1}\right),
\label{eq:ham0}
\end{align}
where $\hat{c}_{i,\sigma}^{\dagger}$ ($\hat{c}_{i,\sigma}$) creates (annihilates) a fermion with spin $\sigma=\uparrow,\downarrow$ at site $i$. Here $\vec{S}_i=\frac{1}{2}\sum_{\alpha\beta}\hat{c}_{i,\alpha}^{\dagger} \vec{\sigma}_{\alpha,\beta}\hat{c}_{i,\beta}$ is the spin-1/2 operator with Pauli matrices $\vec{\sigma}$, and $\hat{n}_i=\sum_{\sigma}\hat{c}_{i,\sigma}^\dagger \hat{c}_{i,\sigma}$ is the density operator. The local modulation is introduced by the site-dependent offset $\delta t_i$ on the hopping for the local site $i$ as
\begin{align}
  \delta t_{i} & =\lambda\cos\left(2\pi i/p+\phi\right).
  \label{eq:modulation}
\end{align}
Here $\lambda$ is the amplitude of the modulation, $p$ is the size of the unit cell, and $\phi$ is the phase factor. These kinds of modulations are very common in researching topological insulators in 1D optical superlattices\cite{Lang2012,Hu2014,Guo2016,Stenzel2019,Hu2019}. Taking into account the $t$-$J$ model is derived from the Hubbard model in the large $U$ limit, we restrict values of parameters to keep $J=4t^{2}/U$ everywhere in real space and therefore set local parameters as 
\begin{align}
  t_{i} & = -t+\delta t_{i}, \nonumber \\
  J_{i} & = Jt^2_{i}/t^2.
\end{align}
To avoid possible ground-state localization~\cite{Aubry1980,Sokoloff1980}, we consider the commensurate lattice with integer $p$. Specifically, we focus on $p=4$ in this work, and other finite even values behave similar phenomena. For convenience, we set $t=1$ as the energy unit and fixed the modulation amplitude $\lambda=0.2$. The model is then systematically investigated at different particle fillings $\rho$ and exchange interaction $J$.

The system described by Eq.~(\ref{eq:ham0}) has $U(1)$ symmetry with conserved total particle number $N_\sigma=\sum_{i}\hat{n}_{i,\sigma}$ for spin species $\sigma$. In practice, we employ the density-matrix renormalization group (DMRG) method~\cite{White1992,White1993} and numerically compute the $m$th eigenstate $\Psi_m(L,N_{\uparrow},N_{\downarrow})$ of energy $E_m(L,N_{\uparrow},N_{\downarrow})$ with fixed good quantum numbers. For convenience, we adopt shortened forms $\Psi_m(N_{\uparrow}, N_{\downarrow})$ and $E_m(N_{\uparrow}, N_{\downarrow})$ when the information of the system size $L$ is clear. 
We dynamically~\cite{Legeza2003} adopt DMRG many-body states to make sure the maximum truncation error is of the order of $10^{-8}$. The number of DMRG kept states is up to 1000 in most calculations but can be very large in some rare cases, for example, up to 8000 in computing the first excited state with degeneracy. 
Open boundary conditions (OBCs) and periodic boundary conditions (PBCs) are used in different calculations for different purposes. Conclusions are made according to the extrapolated results to the thermodynamic limit. 

\subsection{Energy criteria for bulk state}

We adopt several energy criteria to characterize the phase and the ground-state phase diagram. The conducting and insulating bulk state can be determined by the charge excitation gap
\begin{align}
  \Delta E_{C}=&[E_0(N_\uparrow+1,N_\downarrow+1)+E_0(N_\uparrow-1,N_\downarrow-1)\nonumber \\
  &-2E_0(N_\uparrow,N_\downarrow)]/2,
\end{align}
and the zero (nonzero) charge gap indicates the continuous (discrete) charge excitation. The spin gap is defined as the excitation energy from a singlet to a triplet state
\begin{align}
  \Delta E_{S}=E_0(N_\uparrow+1,N_\downarrow-1) -E_0(N_\uparrow,N_\downarrow),
\end{align}
which distinguishes the gapped and gapless spin excitation. Another important energetic criterion is the binding energy 
\begin{align}
\Delta E_{B}=&E_0(N_\uparrow+1,N_\downarrow+1)+E_0(N_\uparrow,N_\downarrow) \nonumber \\
&-2E_0(N_\uparrow+1,N_\downarrow),
\end{align}
which compares the energy of two interacting particles (or holes, depending on the filling) with that of two noninteracting ones, and $\Delta E_{B}<0$ indicates a tendency toward pair formation. Note that all these energy criteria are meaningful only in the thermodynamic limit.

\subsection{Correlation functions and structure factors}

To further characterize different phases, we compute correlation functions and correspondence structure factors. Despite this a quasi-periodic system, we adopt correlations between sites instead of unit cells since two essentially give the same physics~\cite{Reja2016} (see Appendix~\ref{appendix:A} for more information). With the correlation function in a generic form $X_{ij}$ between site $i$ and $j$, one can extract the correlation decay 
\begin{align}
\label{eq:Xr}
  X(r) = \frac{1}{\cal N}\sum_{|i-j|=r}X_{ij},
\end{align}
and the structure factor 
\begin{align}
\label{eq:Xk}
  X(k) = \frac{1}{L}\sum_{i,j=1}^{L}\mathrm{e}^{\mathrm{i}k(i-j)} X_{ij}.
\end{align}
Here $\cal N$ is the total number of pairs $\{i,j\}$ satisfying
$|i-j|=r$, and $k$ is the momentum. The superconducting order can be captured by the singlet-pairing correlation function defined as
\begin{align}
  P_{ij}^S = \langle \Delta^\dagger_i \Delta_j \rangle,
\end{align}
where 
\begin{align}
  \Delta_i = \frac{1}{\sqrt{2}} \left(\hat{c}_{i,\uparrow}\hat{c}_{i+1,\downarrow} - \hat{c}_{i,\downarrow}\hat{c}_{i+1,\uparrow} \right) 
\end{align}
is the annihilation operator for the singlet pair on two nearest sites. We are also interested in the density-density correlation function
\begin{align}
  N_{ij} = \langle \hat{n}_i \hat{n}_j \rangle - \langle \hat{n}_i \rangle \langle \hat{n}_j \rangle,
\end{align}
the spin-spin correlation function
\begin{align}
  S_{ij} = \langle \hat{S}^z_i \hat{S}^z_j \rangle,
\end{align}
and the single-particle Green's function
\begin{align}
  G^\sigma_{ij} = \langle \hat{c}_{i,\sigma} \hat{c}^\dagger_{j,\sigma} \rangle.
\end{align}

\subsection{topological invariant}

In the gapped state, we characterize the nontrivial topology by the Chern number
\begin{align}
\label{eq:Chern}
C = \frac{1}{2\pi}\int_0^{2\pi}d\phi\int_0^{2\pi}d\theta~F(\phi,\theta)
\end{align}
defined in the 2D parameter space ($\phi,\theta$), where $F(\phi,\theta)=\mathrm{Im}\left (\langle \frac{\partial \Psi}{\partial \phi}|\frac{\partial \Psi}{\partial \theta}\rangle - \langle \frac{\partial \Psi}{\partial \theta}|\frac{\partial \Psi}{\partial \phi}\rangle \right)$ is the Berry curvature, and $\theta$ is the phase factor for twisted boundary conditions~\cite{Niu1985,Xiao2010}. Specifically, we impose twisted boundary conditions via the replacement $\hat{c}_{j,\sigma} \rightarrow e^{\mathrm{i}\delta\theta j}\hat{c}_{j,\sigma}$, where $\delta\theta = \theta/L$ is the phase gradient. We compute the Chern number for insulating states using the method in Ref.~\cite{Fukui2005,Varney2011} on a $12\times12$ discrete grid, using both exact wavefunctions from the exact diagonalization method for smaller system size ($L=16$) and matrix-product-state wavefunctions from DMRG for a larger one ($L=64$). These two methods give the same result.

\section{Ground-state phase diagram and Enhanced superconductivity}
\label{sec:bulk}

\begin{figure}[!t] 
  \includegraphics[width=1\columnwidth]{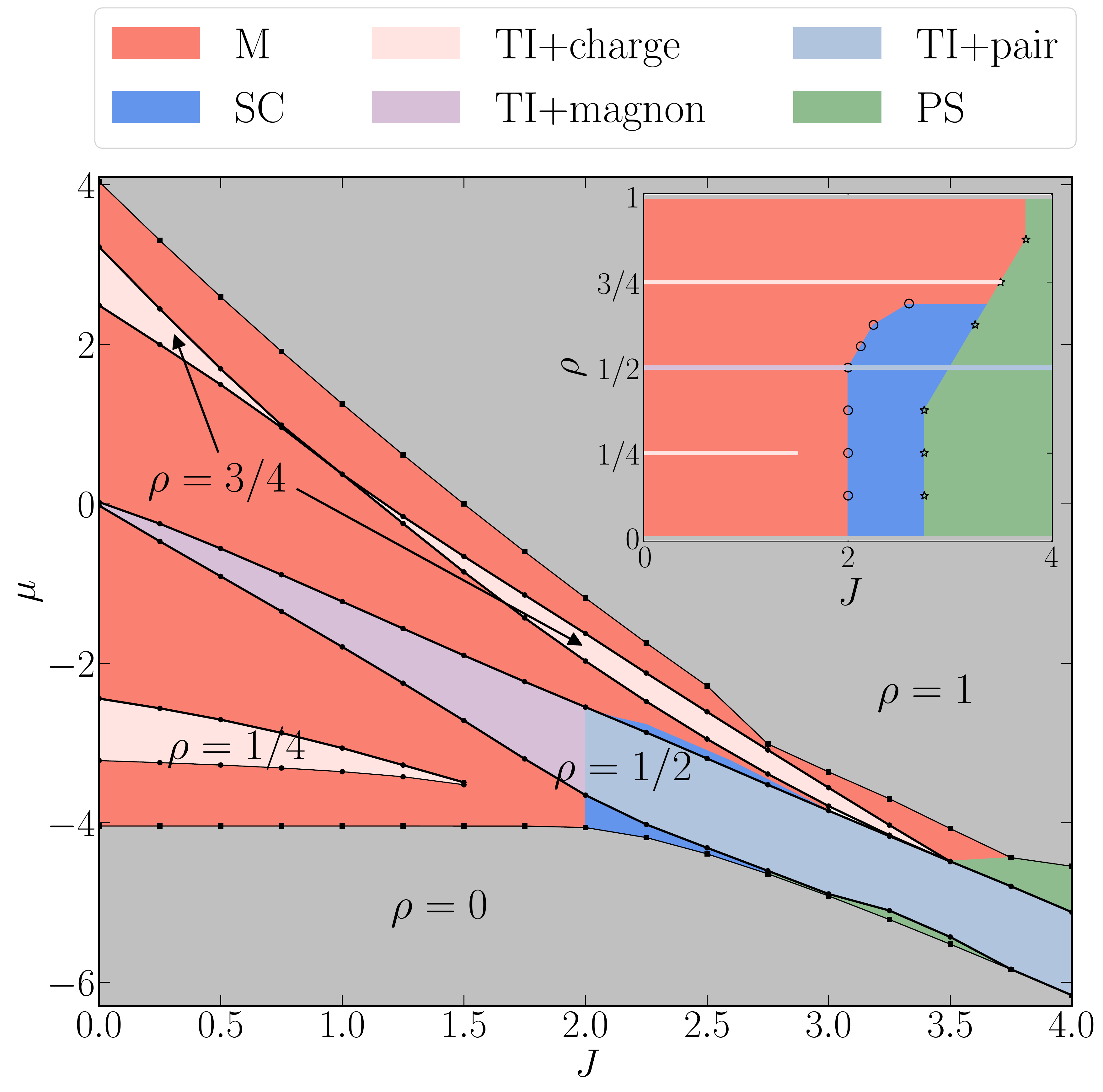}
  \caption{The ground-state phase diagram consists of the metallic phase (M), superconducting phase (SC), phase separation (PS), and different topological insulating regions (TI). For the topological insulating regions, the gapless quasi-particle excitation is labeled as well. The gray region of $\rho=0$ ($\rho=1$) denotes the particle empty (full) state. The phase diagram is drawn in the $\mu$-$J$ ($\rho$-$J$) plane in the main (inset) panel. Here we use $\phi=0$ for all calculations and results are extrapolated to the thermodynamic limit. }
  \label{fig:phase_diagram}
\end{figure}

\begin{figure*}[!t] 
  \includegraphics[width=1\linewidth]{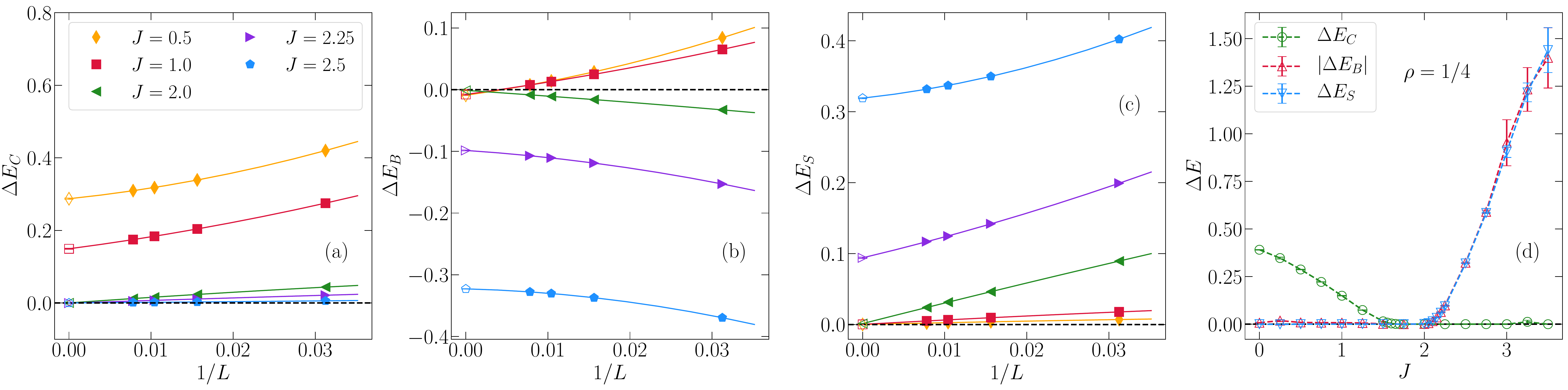}
  \caption{Energy criteria for metallic, superconducting, and insulating phases: (a) the charge gap, (b) the binding energy, and (c) the spin gap as functions of $1/L$ for different $J$s, with filled symbols for results in the finite systems and empty symbols for the extrapolated result in the limit $1/L\rightarrow 0$. (d) Extrapolated results in the thermodynamic limit as a function of $J$.  Here we use data of $L=32$, $64$, $96$, and $128$ with OBCs, at the filling $\rho=1/4$. Panels (a-c) share the same legend. Here and after, error bars for extrapolated results denote the uncertainty from the least square polynomial fitting. }
  \label{fig:energy_rho2o8}
\end{figure*}

We summarize our main result by the ground-state phase diagram in Fig.~\ref{fig:phase_diagram}, which contains the metallic phase, the superconducting phase, the phase separation, and different topological insulating states. To emphasis the gapped phases, in the main panel, we plot the phase diagram in the grand canonical ensemble in the presence of a uniform chemical potential $\mu$. The ground-state energy of the grand ensemble Hamiltonian
\begin{align}
  {\cal H}_\mu = {\cal H} + \mu\sum_i\hat{n}_i
  \label{eq:ham1}
\end{align}
can be easily obtained by $E_0(N_{\uparrow}, N_{\downarrow},\mu) = E_0(N_{\uparrow}, N_{\downarrow}) + \mu N$, with $N = \sum_\sigma N_\sigma$ is the total number of particles. Insulating phases ($\Delta E_C>0$) lie in commensurate fillings, i.e., $\rho=1/4$, $1/2$, and $3/4$, where the particle filling is defined as $\rho=N/L$. Boundaries between the metallic and superconducting state can be obtained from the binding energy $\Delta E_B$. The negative inverse compressibility determines the boundary entering the phase separation (see Appdendix~\ref{appendix:B}). In this work, we restrict ourselves to the case with $N_\uparrow = N_\downarrow$ and $\rho \in (0,1)$. The system at $\rho=1$, where the Hamiltonian in Eq.~(\ref{eq:ham0}) degenerates to a quasi-periodic Heisenberg chain, which also shows nontrivial topological properties and has been investigated in previous literatures~\cite{Hu2014,Hu2015}. In this section, we focus on the bulk state characterization with $\phi=0$ in all calculations, and the topological nature of insulating states is discussed in Sec.~\ref{sec:edge}. 

More specifically, we display in Fig.~\ref{fig:energy_rho2o8} energy criteria for bulk phases by taking the filling $\rho=1/4$ as an example. The finite size value of the charge gap $\Delta E_C$ and its extrapolation by a second-order polynomial fitting is shown in Fig.~\ref{fig:energy_rho2o8}(a). In the thermodynamic limit, there exists a finite charge gap at small $J$s, and the gap closes for larger values of $J$. This indicates a transition from the charge insulating to the conducting state. Similarly, the finite size extrapolation of the binding energy $\Delta E_B$ and the spin gap $\Delta E_S$ is shown in Fig.~\ref{fig:energy_rho2o8}(b) and (c) respectively. By plotting these energy criteria together in Fig.~\ref{fig:energy_rho2o8}(d), one can tell bulk phases and transitions at this filling immediately. At small $J$s, the system behaves as a charge insulating state with the gapless spin excitation. At intermediate exchange interactions, the system is in the conducting state of unpaired fermions ($\Delta E_B = 0$), dubbed the metallic phase. At larger $J$s, while the charge excitation remains gapless, significant negative binding energy occurs. In this conducting state of paired fermions, the binding energy $\Delta E_B$ and the spin gap $\Delta E_S$ have the same magnitude, showing evidence for the singlet-paired superconductivity. Compared to the homogeneous $t$-$J$ chain, the spin gap is greatly enhanced in the modulated setting. For example, while we have $\Delta E_S \approx 0.32$ at $J=2.5$ in Fig.~\ref{fig:energy_rho2o8}(d), the spin gap is around $0.01$ for the same $J$ and $\rho$ in the standard $t$-$J$ model~\cite{Moreno2011}.

\begin{figure}[!t] 
  \includegraphics[width=1\columnwidth]{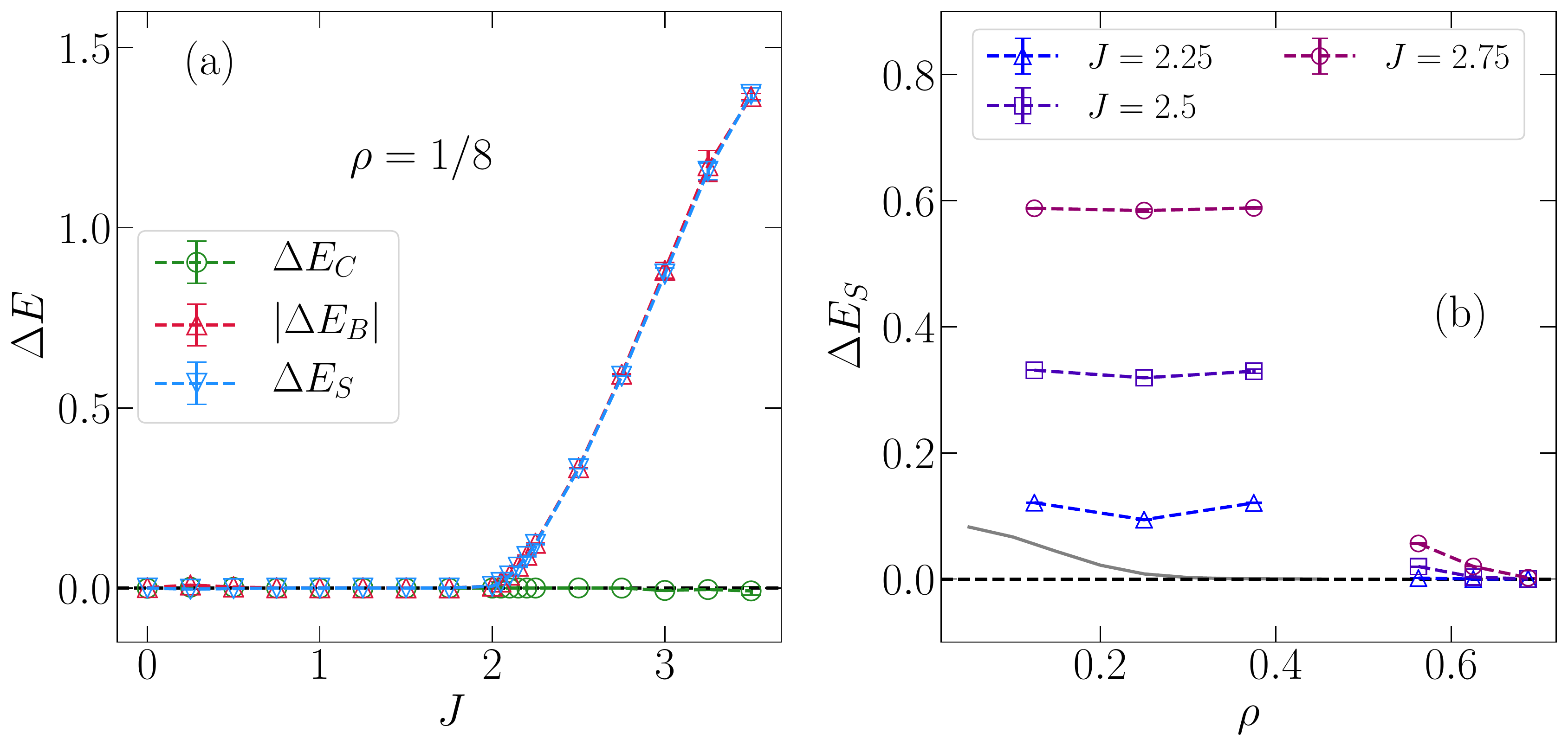}
  \caption{(a) Energy criteria as a function of $J$ at $\rho=1/8$. (b) Spin gap $\Delta E_S$ as a function of $\rho$ for typical $J$s. Extrapolated results in the thermodynamic limit are obtained from data of $L=32$, $64$, $96$, and $128$ with OBCs. Gary solid line in (b) denotes the spin gap at $J=2.5$ for the standard $t$-$J$ chain with data extracted from Ref.~\cite{Moreno2011}. }
  \label{fig:energy_diffrho-1}
\end{figure}

At incommensurate fillings ($p\times\rho$ is not an integer), the system is always in the conducting phase (before phase separation) with gapless charge excitation. Taking $\rho=1/8$ in Fig.~\ref{fig:energy_diffrho-1}(a) as an example, there is a transition from the metallic state to the superconducting state with singlet pairs. We further display the spin gap versus density $\rho$ in Fig.~\ref{fig:energy_diffrho-1}(b) in the superconducting state and find that the spin gap for the same $J$ almost does not decay as particle density $\rho$ increases for $\rho<1/2$. This phenomenon is in sharp contrast to the homogeneous case, where $\Delta E_S$ dies off very rapidly, as denoted by the gray solid line in Fig.~\ref{fig:energy_diffrho-1}(b). The strongly enhanced superconductivity in this work lies in the very flat band structure in the ground-state phase diagram~\ref{fig:phase_diagram}, which is in agreement with the argument that the fermionic pairing in flat bands would lead to more robust pairs and higher
critical temperatures~\cite{Heikkila2016,Mondaini2018}. However, at fillings $\rho>1/2$, the spin gap becomes less and less significant and finally disappears around $\rho\approx 11/16$. Present results for the generally modulated $t$-$J$ chain do not support superconductivity at physical-relevant parameters in materials (low-doping and very small $J$), which is different from what was found in $t$-$J$ segments~\cite{Reja2016}. 

\begin{figure}[!t] 
  \includegraphics[width=1\columnwidth]{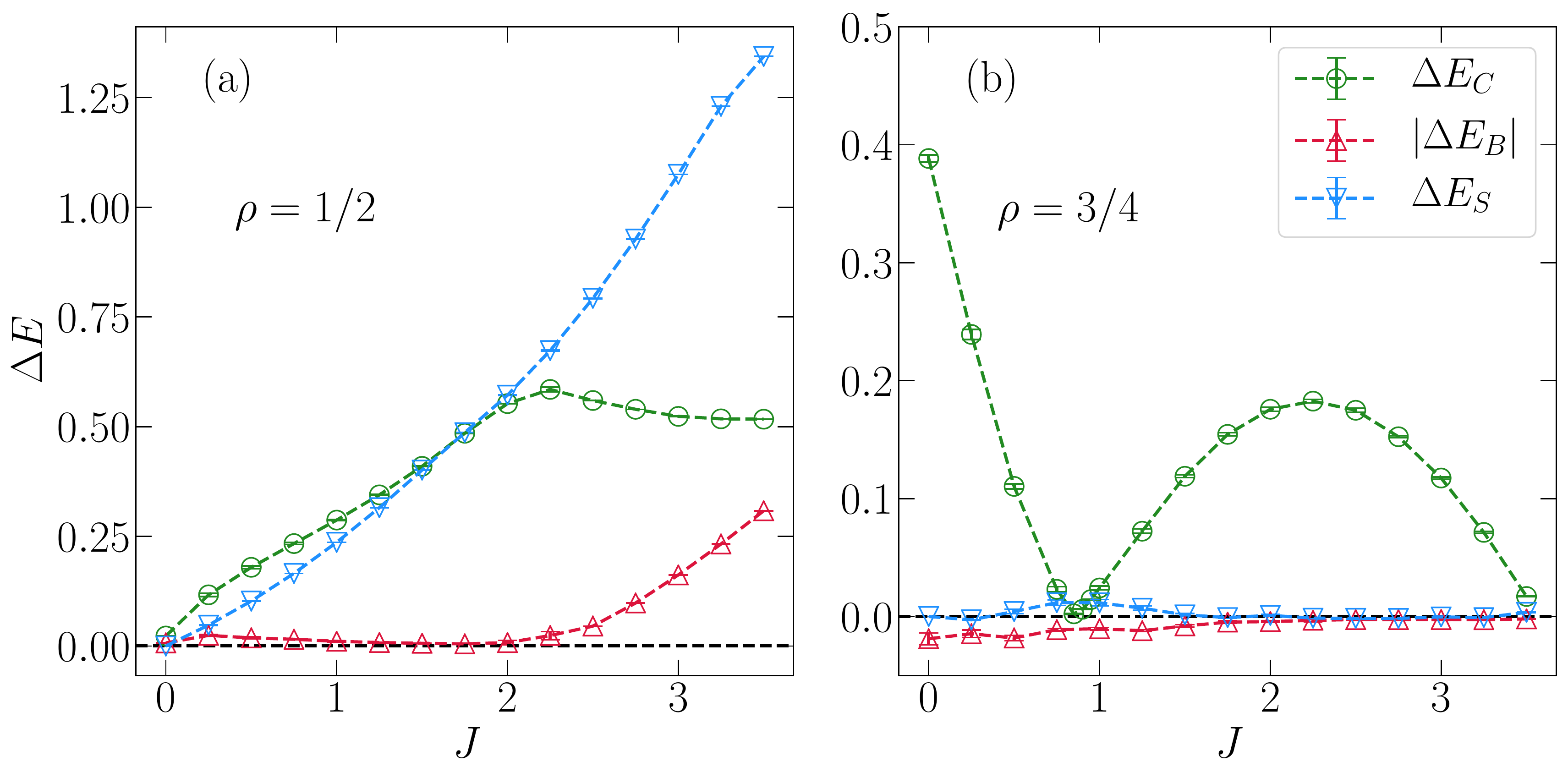}
  \caption{Energy criteria for different typical fillings in the thermodynamic limit at (a) $\rho=1/2$ and (b) $\rho=3/4$. Extrapolated results in the thermodynamic limit are obtained from data of $L=32$, $64$, $96$ and $128$ with OBCs for $\rho=1/2$, and $L=32$, $48$, $64$, $80$, $96$ with PBCs for $\rho=3/4$. All panels share the same legend.}
  \label{fig:energy_diffrho-2}
\end{figure}

At commensurate fillings, one can observe apparent charge gaps, but corresponding insulating states show different features. At $\rho=1/4$, the insulating state at small $J$s has gapless spin excitation and no pair formation, as discussed in the previous text and shown in Fig.~\ref{fig:energy_rho2o8}. At $\rho=1/2$, the spin gap is always nonzero, and $\Delta E_B$ is finite only at $J>2$, as shown in Fig.~\ref{fig:energy_diffrho-2}(a). Things are more complicated for $\rho=3/4$ in Fig.~\ref{fig:energy_diffrho-2}(b), where appear two disconnected insulating states with a gap-closing point around $J\approx 0.85$. At this special filling with $J>0.85$, 
the energy criteria using OBCs and PBCs are different at $\phi=0$. In contrast, different boundary conditions give the same result in the thermodynamic limit in all other fillings. Therefore, we adopt PBCs for the ground-state phase diagram at $\rho=3/4$ and explain the underlying physics in Sec.~\ref{sec:insu3}. Different bulk properties in these different insulating states essentially affect their topological nature and edge excitations, which will be discussed in detail in the next section. 

\subsection{Density-density correlations and the Luttinger parameter}

\begin{figure}[!t] 
  \includegraphics[width=0.9 \columnwidth]{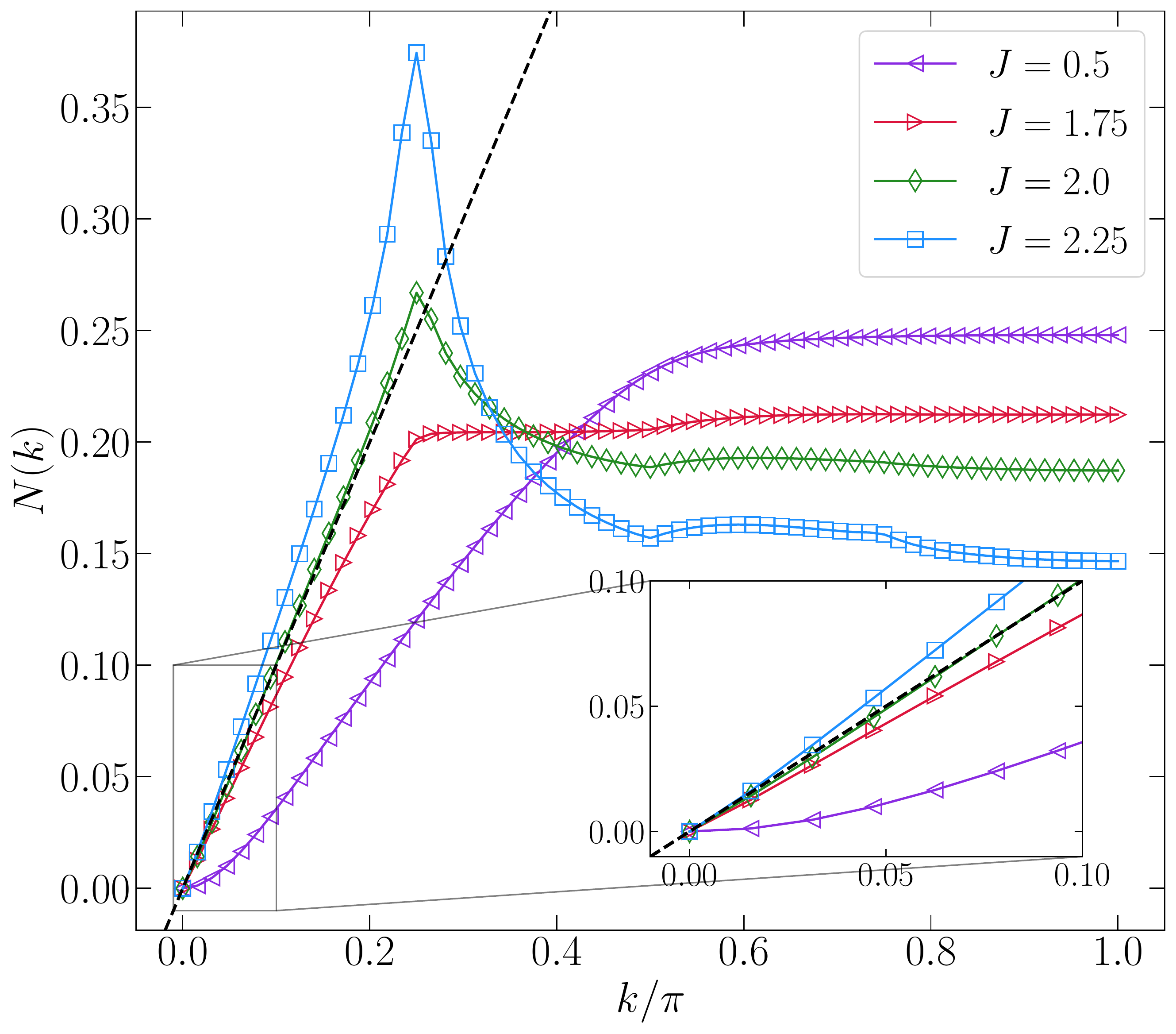}
  \caption{Structure factor $N(k)$ at $\rho=1/4$. The black dashed line denotes $N(k)/(k/\pi)=1$. Here $L=160$ and OBCs are adopted.}
  \label{fig:Nk}
\end{figure}

In this subsection, we follow the standard way of the Luttiger liquid process to review bulk phases by analyzing the density-density correlation function and the Luttiger parameter~\cite{Giamarchibook,Moreno2011}. The latter is confirmed to be valid for quasi-periodic lattices even using correlations between sites~\cite{Valencia2002,Reja2016}. The Luttiger parameter can be determined by the structure factor $N(k)$ at the small momentum $k$, as
\begin{align}
  K_\rho = N(k)/(k/\pi) , {k\rightarrow 0},
\end{align}
where $N(k)$ is the structure factor of the density-density correlation function, as shown in Fig.~\ref{fig:Nk}, we display $N(k)$ at the filling $\rho=1/4$, where one can see metallic, superconducting, and insulating phases. At a small interaction $J=0.5$, where the system is in the insulating phase and the charge excitation is gapped (see Fig.~\ref{fig:energy_rho2o8}), $N(k)$ shows quadratic behavior at small momentum $k$. In both gapless phases for charge excitations, $N(k)$ has a clear linear behavior at small momentum, and $K_\rho$ is smaller (larger) than $1$ in the metallic (superconducting) phase. At the critical point $J=2$, the Luttiger parameter $K_\rho$ is close to 1. Results from the structure factor $N(k)$ and the Luttiger parameter agree with the ground-state phase diagram by energy criteria. Note that Fig.~\ref{fig:Nk} shows results for a finite system with $L=160$. Although we believe that structure factors and energy criteria must give the same conclusion in the thermodynamic limit, one should pay careful attention to the finite size effect. 

\subsection{Pairing correlations and the structure factor}

\begin{figure}[!ht] 
  \includegraphics[width=0.9 \columnwidth]{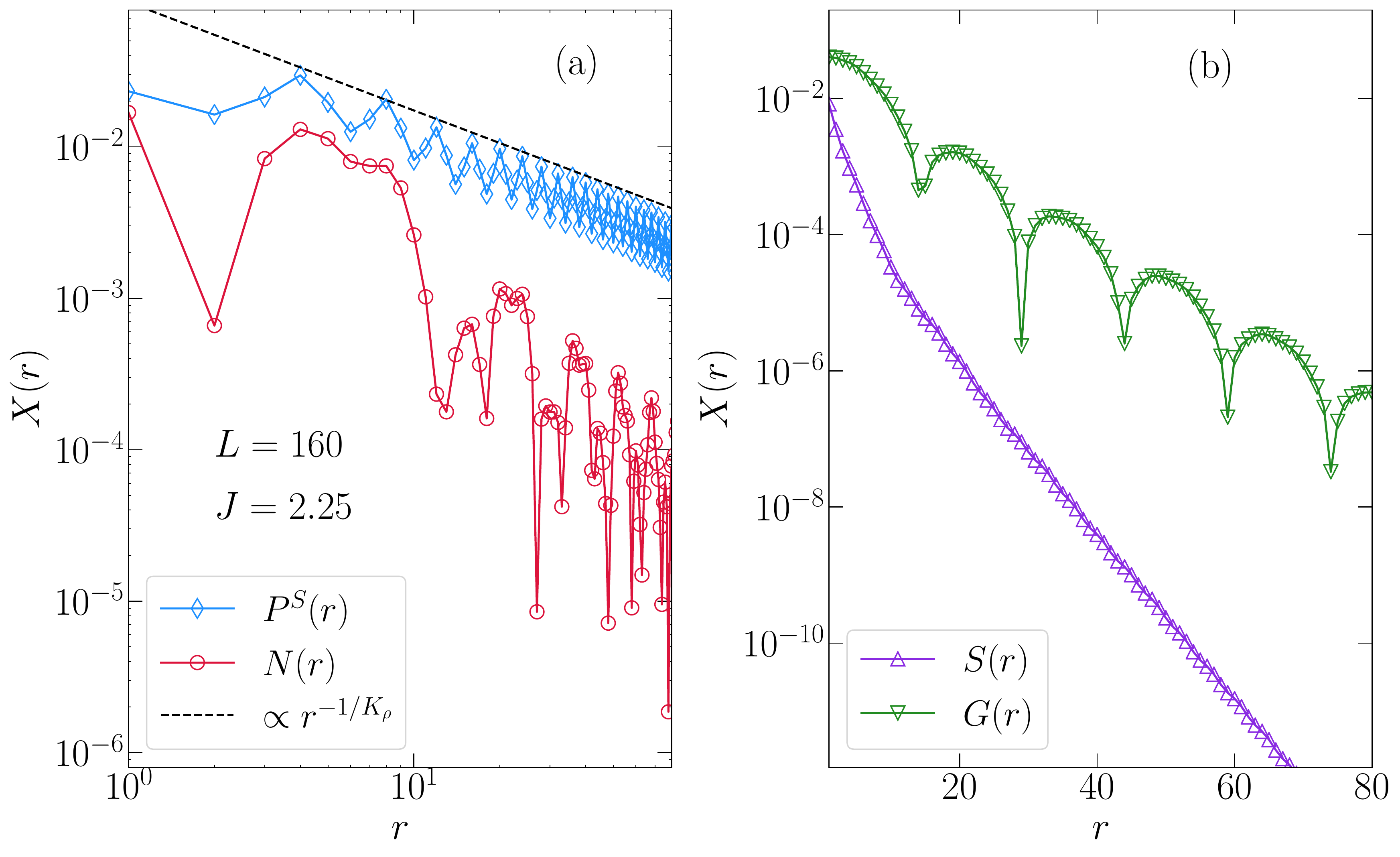}
  \caption{ Correlation functions in the superconducting states at $J=2.25$ and $\rho=1/8$: (a) Density-density and pairing correlation functions in the log-log scale, and (b) spin-spin correlation function and single-particle Green's function in the linear-log scale. The black dashed line in (a) depicts the power-law fit of the pairing correlation function, with $K_\rho=1.4$ for this finite size. Here $L=160$ and OBCs are adopted.}
  \label{fig:corr1}
\end{figure}

In the ladder geometry, the $t$-$J$ Hamiltonian can support superconductivity with the significant spin gap, negative binding energy, and the dominant singlet pairing correlation~\cite{Hayward1995,Cheng2018}. 
While in this subsection, we focus on the correlations and the relevant elementary excitations in the superconducting phase. As shown in Fig.~\ref{fig:corr1}, we display correlations at $\rho=1/8$ as an example. While both the pairing and density-density correlation functions have the power-law decay, $P^S(r)$ decays much slower than $N(r)$. The spin-spin correlation and the single-particle Green's function,
which correspond to spin and fermion excitations, both decay exponentially. This result agrees with the occurrence of the spin gap and negative binding energy in the superconducting state.  

\begin{figure}[!t] 
 \includegraphics[width=0.9 \columnwidth]{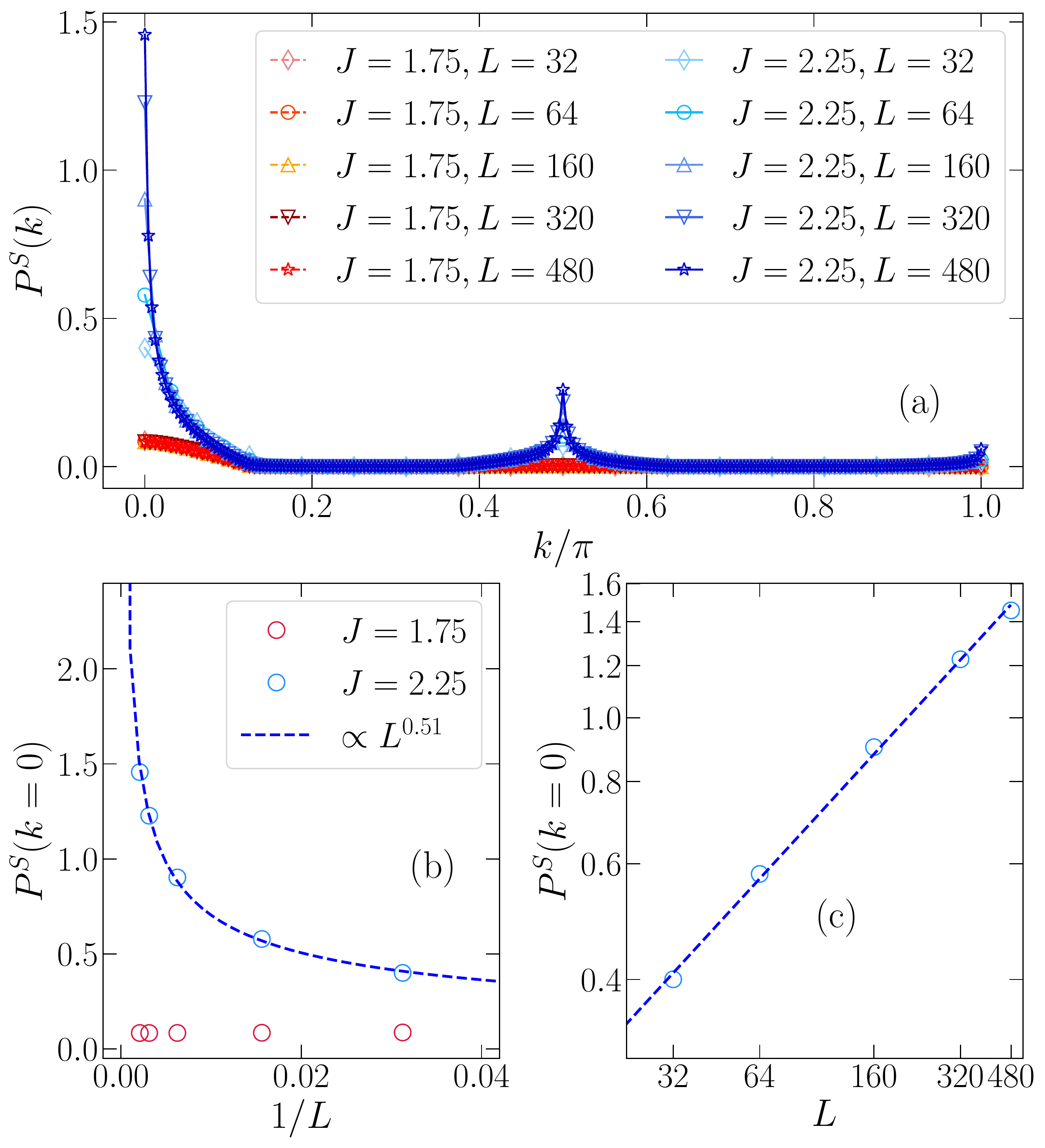}
 \caption{(a) Structure factor $P^S(k)$ of singlet pairing correlations as a function of momentum $k$ in the superconducting ($J=2.25$) and metallic ($J=1.75$) state at $\rho=1/8$ for different $L$. The system size dependence of $P^S(k=0)$ is shown in (b) and (c), where the blue dashed line depicts the power-law fit. Panels (b) and (c) share the same legend.}
 \label{fig:PSk1}
\end{figure}

In the spin-gapped superconducting state, the pairing structure factor $P^S(k)$ at zero momentum was shown to be diverging in the thermodynamic limit ~\cite{Moreno2011}. Here we present $P^S(k)$ in Fig.~\ref{fig:PSk1}(a) as a function of $k$, in both the metallic and superconducting phase for comparison. In the superconducting phase, the pairing structure factor has a sharp peak at $k=0$, and $P^S(k=0)$ increases as the system size $L$ increases. In contrast, despite a broad maximum at zero momentum, $P^S(k)$ is much smaller in the metallic phase and almost independent of the system size. More Specifically, by displaying the size dependence of $P^S(k=0)$ explicitly in Fig.~\ref{fig:PSk1}(b) and (c), it is clear that $P^S(k=0)$ has a power-law growth as $L$ increases. However, while the number of pairs at zero momentum given by $P^S(k=0)$  diverges, the density of pairs given by $P^S(k=0)/L$ is infinitely small in the thermodynamic limit. In the quasi-periodic $t$-$J$ chain, we only have the quasi-long-range superconducting order, which does not go beyond the ladder geometry. 

\section{insulating states and topological nature}
\label{sec:edge}

The system described by Eq.~(\ref{eq:ham0}) has several insulating regions at commensurate fillings, as discussed in Sec.~\ref{sec:bulk}. In this section, we reinvestigate these gapped phases from the point of view of their topological nature. Before going into details, we emphasize that the interaction plays a critical role in the modulated $t$-$J$ chain. For example, we observe insulating states at $\rho=q\times p$ for any integer $q$ smaller than $p$, while spinful fermions in the noninteracting setting show insulating behaviors only at even $q$s, in the absence of the external Zeeman field. In all insulating regions, our calculations give the nonzero topological Chern number. Furthermore, the Chern number is different at different fillings, that is, $C=1$ at $\rho=1/4$, $C=2$ at $\rho=1/2$ and $C=-1$ at $\rho=3/4$. Therefore, in the following, we discuss these topological insulating states at each different filling in one subsection.

\subsection{\texorpdfstring{$\rho=1/4$}{Lg}}
\label{sec:insu1}

The system behaves as charge insulating states at densities $\rho=1/4$ with small exchange $J$s, as shown in the ground-state phase diagram in Fig.~\ref{fig:phase_diagram}. In this region, particles do not form singlet pairs ($\Delta E_B = 0$) and the spin excitation is gapless ($\Delta E_S = 0$). The quasi-particle excitation, which is gapped in bulk but gapless at edges, if it exists, can correspond to the single-particle spin-1/2 fermionic mode. Along this line, we define the single-particle fermionic excitation energy as the ground-state energy difference by adding a single fermion 
\begin{align}
  \delta E_f = E_0(N_\uparrow,N_\downarrow+1) - E_0(N_\uparrow,N_\downarrow),
\end{align}
and the corresponding onsite charge (spin) excitation $\delta_f \langle \hat{n}_i\rangle$ ($\delta_f \langle \hat{S}^z_i\rangle$) can be defined as differences between these two ground states. Here we consider adding a spin-down fermion, and the case for adding a spin-up one is symmetric and obvious. 

\begin{figure}[!t] 
  \includegraphics[width=1\columnwidth]{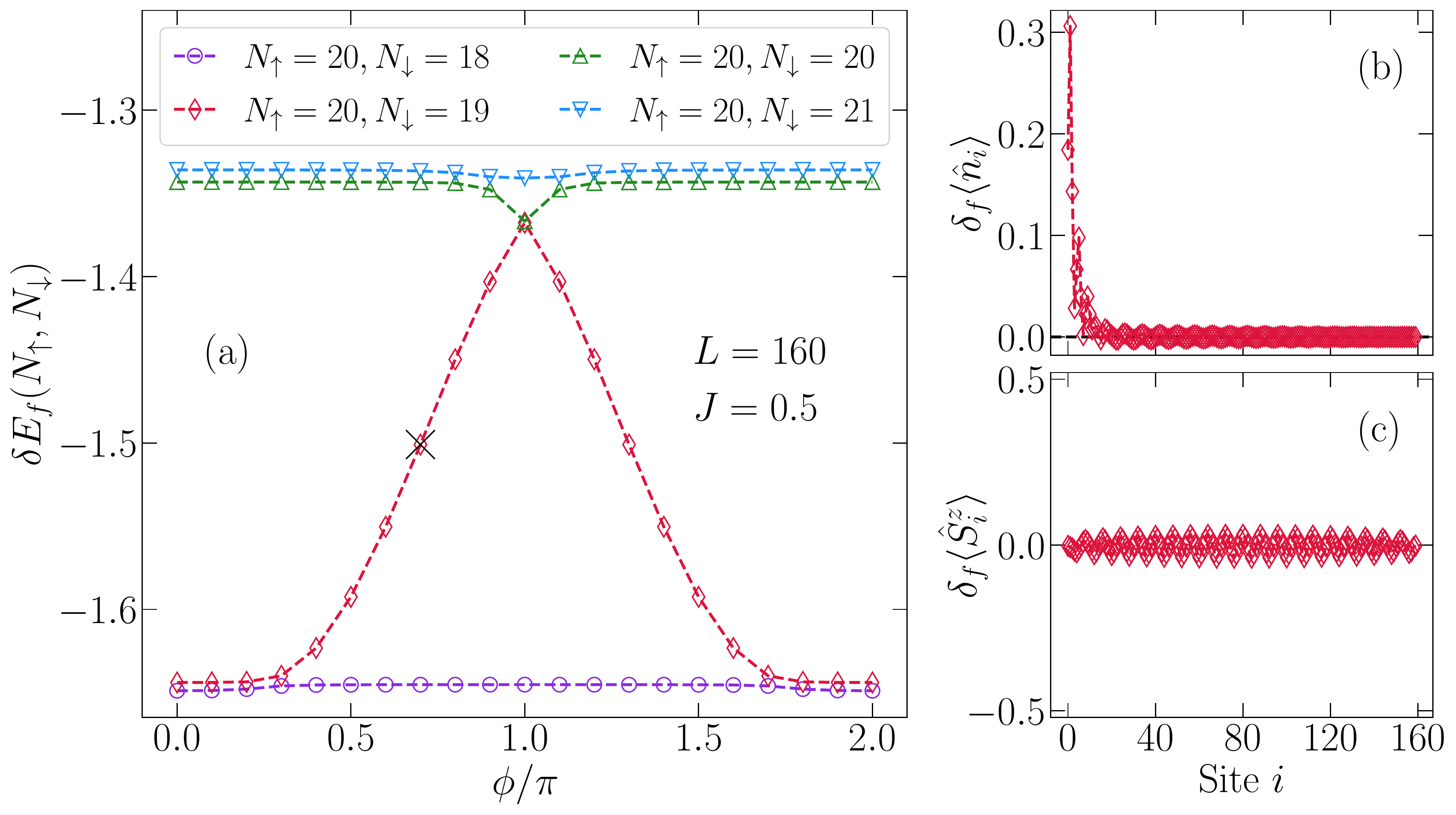}
  \caption{ (a) The quasi-particle (single fermion) spectrum as a function of the phase factor $\phi$ near the filling $\rho=1/4$. (b) The corresponding onsite charge and spin difference for the two adjacent many-body ground states, at parameters labeled by the black cross in (a). Here results are from calculations of the system with $J=0.5$, $L=160$ and OBCs. All panels share the same legend. }
  \label{fig:topo_1}
\end{figure}

As a routine in characterizing the bulk-edge correspondence for 1D topological insulators, we display the quasi-particle spectrum as a function of the phase factor $\phi$ in Fig.~\ref{fig:topo_1}(a), in which we show four typical quasi-particle levels and nearby lower (upper) levels with smaller (larger) $N_\downarrow$ are in the continuous band. While there is a significant gap at $\phi=0$, the top level in the lower band lifts as $\phi$ evolves and meets the upper band at $\phi=\pi$. In Fig.~\ref{fig:topo_1}(b) and (c), we plot the low-energy charge and spin excitations, respectively, by taking the in-gap state at $0.7\pi$ as an example. The well-localized charge accumulation at one end of the chain can be observed, but the spin excitation shows no edge-related anomaly. This strong spin-charge separation in the bulk-edge correspondence is essentially different from the noninteracting case, where there exists no quasi-particle that carries one charge but no spin. In this strongly interacted insulating region, the spin-1/2 single fermion contributes the energy for the band touching. However, only the excitation of charge degree of freedom is related to the edge mode, and its spin degree of freedom merges into the bulk.


\subsection{\texorpdfstring{$\rho=1/2$}{Lg}}

In the noninteracting case, a spinless-fermion chain with quasi-periodic modulations described in Eq.~(\ref{eq:modulation}) has insulating states at commensurate fillings $\rho=q/p$ with an integer $q<p$~\cite{Lang2012,Kraus2012}, which results in naive charge and spin gaps (nonzero $\Delta E_C$ and $\Delta E_S$) at $\rho=2q/p$ for the noninteracting spinful fermions. 
In the interacting setting, the charge and spin gap survive at $\rho=1/2$ in modulated $t$-$J$ chains, as shown in Fig.~\ref{fig:energy_diffrho-2}(b). However, the exchange interaction $J$ further divides this gapped phase into two regions according to whether singlet-pairs are formed in the ground state. In the following, we demonstrate that the bulk property also affects the topological nature and the gapless edge modes are different in the two regions. Consequently, there is an interaction-driven transition in the topological aspect as well. 

\begin{figure}[!ht] 
  \includegraphics[width=1\columnwidth]{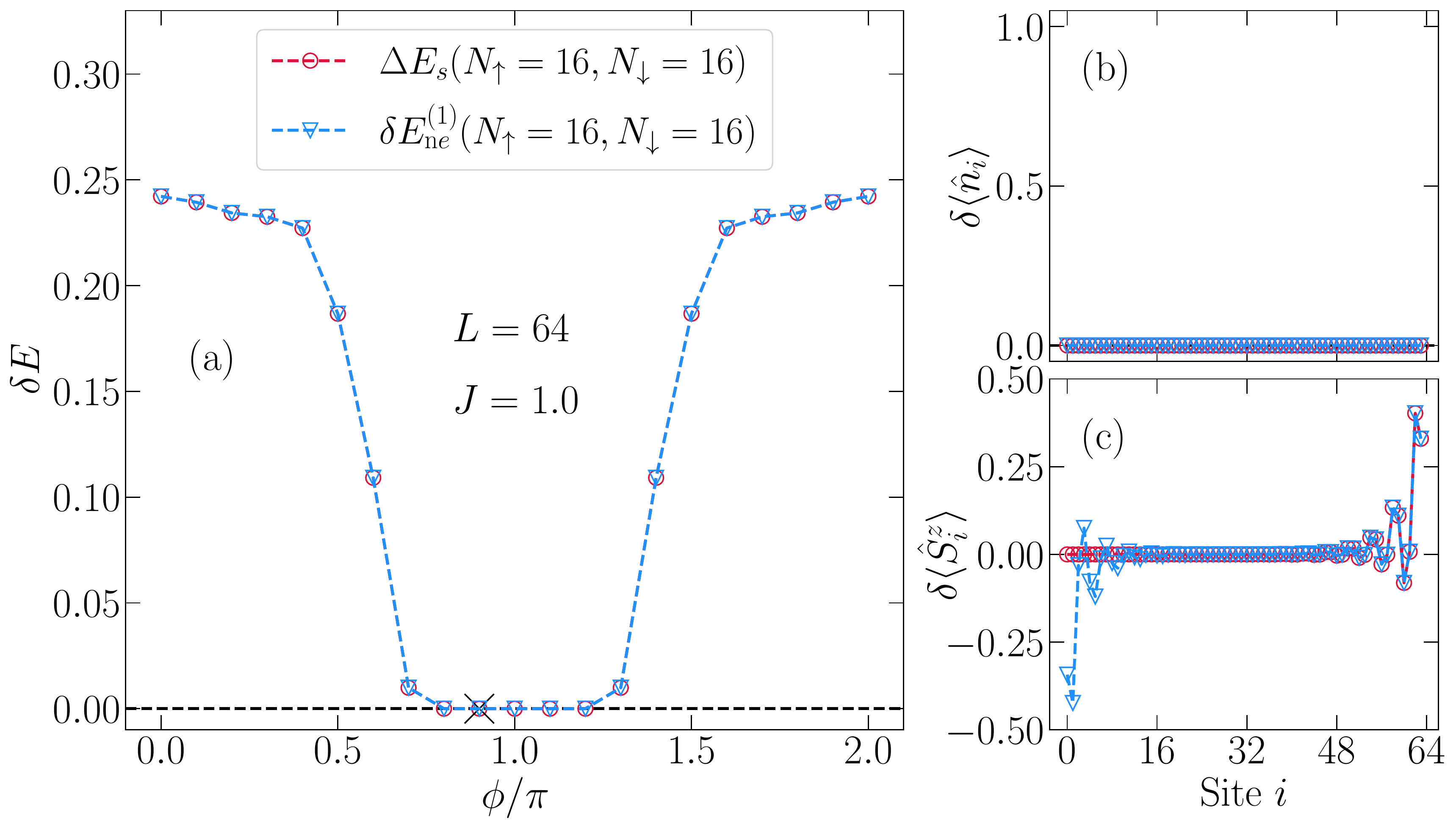}
  \caption{ (a) The spin gap $\Delta E_S$ and the neutral excitation energy $\delta E_{\mathrm ne}^{(1)}$ as a function of $\phi$ at the filling $\rho=1/2$. The corresponding low-energy (b) charge and (c) spin excitations in real space, at parameters labeled by the black cross in (a). Here results are from calculations of the system with $J=1.0$, $L=64$ and OBCs. All panels share the same legend. }
  \label{fig:topo_2a}
\end{figure}

The binding energy is zero at small exchange interactions ($J<2$), with no indication of pair formation. If repeating procedures in Sec.~\ref{sec:insu1} by considering the single fermion as the gapless edge modes, one cannot find level crossing in the quasi-particle spectrum as in Fig.~\ref{fig:topo_1}(a). In the region with both gapped charge and spin excitations, it is useful to examine the neutral excitation to the first excited state to determine which one is the minimum. In Fig.~\ref{fig:topo_2a}(a), we compare the spin-flip energy and the neutral excitation energy $\delta E^{(1)}_{\mathrm ne} = E_1 - E_0$, and find the surprisingly exact match between them for all values of the phase factor $\phi\in[0,2\pi]$. At $\phi=0$, both excitations are gapped, and the two gaps close around $\phi=\pi$. In Fig.~\ref{fig:topo_2a}, we further display the low-energy charge (spin) excitations in real space, which can be defined as the difference of the corresponding observable between two many-body states. The low-energy charge excitation shows no anomaly for both the spin-flip and the neutral excitation, as shown in Fig.~\ref{fig:topo_2a}(b). In contrast, in Fig.~\ref{fig:topo_2a}(c), one can observe clear spin accumulations at one end of the chain when flipping a spin. Spin accumulation occurs at both ends of the chain for the neutral excitation to keep the total spin in the z-direction as $0$. Moreover, at one end of the chain, where the spin accumulation is observed for the spin-flipping, the spin distribution shows great agreement with the neutral excitations. These results confirm the gapless edge modes related to the magnon excitation, which carries integer spin without charge degree of freedom.

\begin{figure}[!t] 
  \includegraphics[width=1\columnwidth]{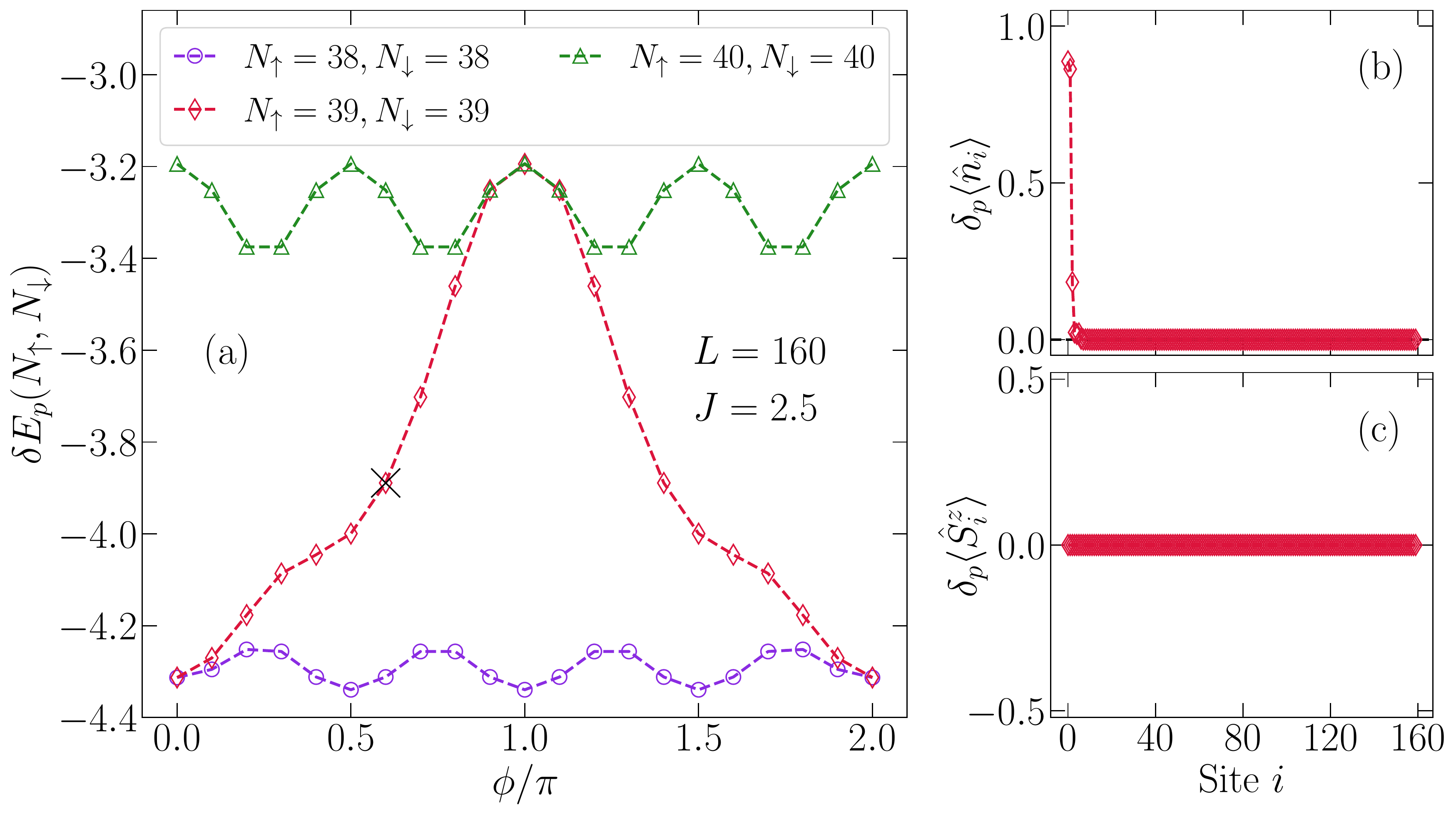}
  \caption{ (a) The quasi-particle (singlet pair) spectrum as a function of the phase factor $\phi$ near the filling $\rho=1/2$. (b) The corresponding onsite charge and spin difference for the two adjacent many-body ground states, at parameters labeled by the black cross in (a). Here results are from calculations of the system with $J=2.5$, $L=160$ and OBCs. All panels share the same legend.}
  \label{fig:topo_2b}
\end{figure}

\begin{figure}[!b] 
  \includegraphics[width=0.85\columnwidth]{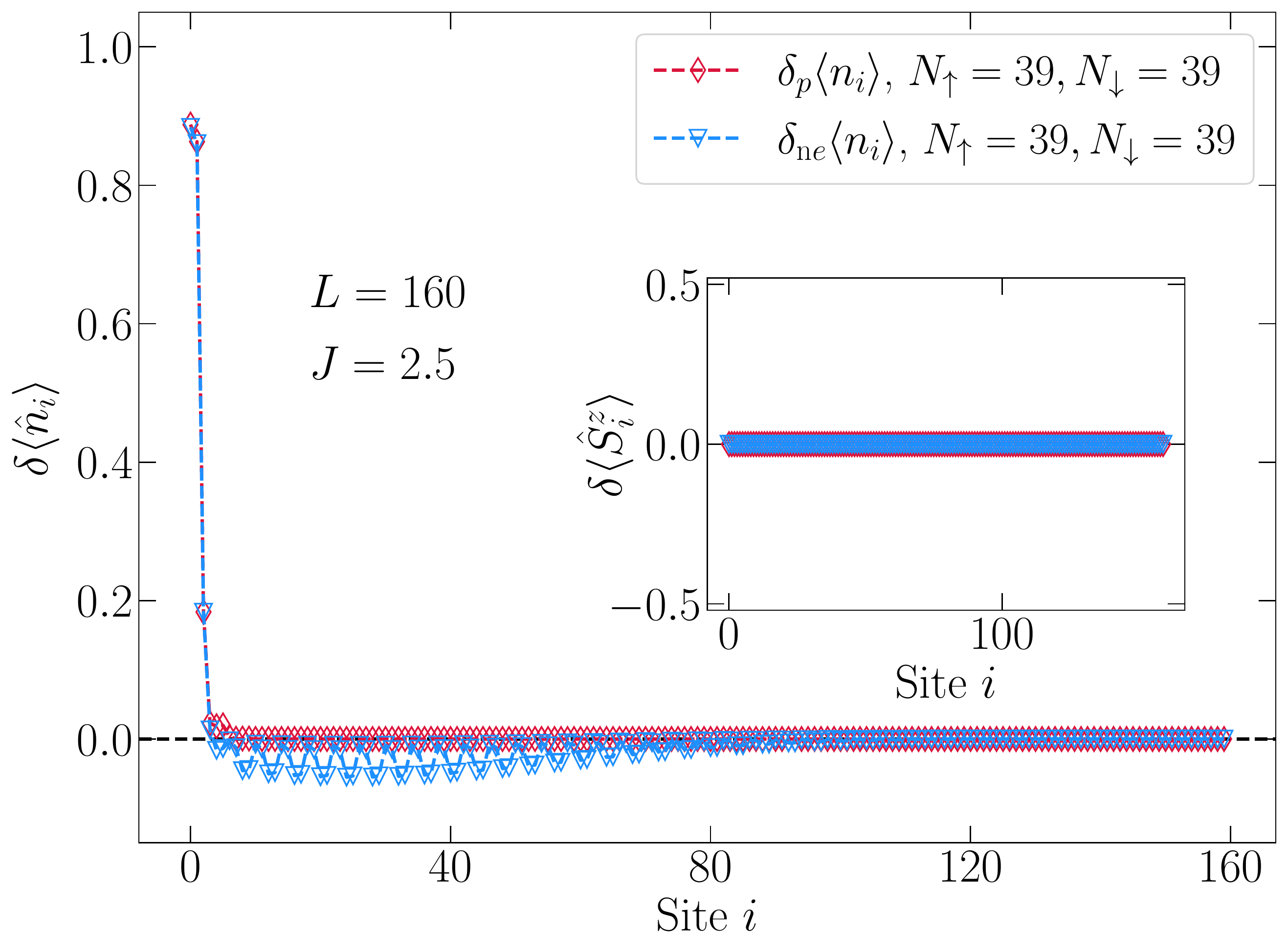}
  \caption{ The low-energy charge (main panel) and spin (inset) excitations in real space, at parameters labeled by the black cross in Fig.~\ref{fig:topo_2b}(a). Here results are from calculations of the system with $J=2.5$, $L=160$ and OBCs.}
  \label{fig:topo_2b1}
\end{figure}

When the interaction $J$ is above a critical $J_c\approx 2$, the negative binding energy occurs, and fermions tend to form singlet pairs. Intuitively, it is interesting to consider the possibility of another kind of bosonic edge mode attributed to singlet pairs. We define the pairing excitation energy
\begin{align}
  \delta E_p = E_0(N_\uparrow+1,N_\downarrow+1) - E_0(N_\uparrow,N_\downarrow),
\end{align}
and present the corresponding quasi-particle spectrum for a typical $J=2.5$ in Fig.~\ref{fig:topo_2b}(a). The quasi-particle spectrum has a considerable gap at $\phi=0$, and the top level of the lower band rises as $\phi$ increases and meets the upper band around $\phi\approx\pi$. Following similar procedures in the previous text, we display the low-energy charge and spin excitations of the in-gap state at $\phi=0.6\pi$ in Fig.~\ref{fig:topo_2b}(b) and (c), respectively. As expected, charge accumulations occur at one end of the chain, and there is no spin anomaly in real space. We further compare the low-energy charge excitation by adding a singlet-pair to which of the neutral excitation in the main panel of Fig.~\ref{fig:topo_2b1}. It is impossible to see the exact match between the two curves at the finite system size since they correspond to excitations with different particle number altering. However, charge accumulations at the end of the chain show great agreement between two kinds of low-energy excitations, and the total charge altering within the peak (roughly six sites) is about 2. Considering that there is no anomaly in the spin channel, one can conclude that singlet-pairs contribute to the gapless edge mode. 

\subsection{\texorpdfstring{$\rho=3/4$}{Lg}}
\label{sec:insu3}

There are two insulating regions at $\rho=3/4$ separated by a gap closing-point around $J_c\approx0.85$, as shown in the ground-state phase diagram~\ref{fig:phase_diagram} and Fig.~\ref{fig:energy_diffrho-2}(d). The spin gap and the binding energy are zero in both regions, which rules out the magnon and singlet pairs as gapless edge modes. When $J$ is smaller than the gap closing point, the bulk-edge correspondence is similar to the case at $\rho=1/4$ as described in Sec.~\ref{sec:insu1}. Therefore, we skip the repeated description and move to the case for larger exchange interactions in the following. 

\begin{figure}[!t] 
  \includegraphics[width=1\columnwidth]{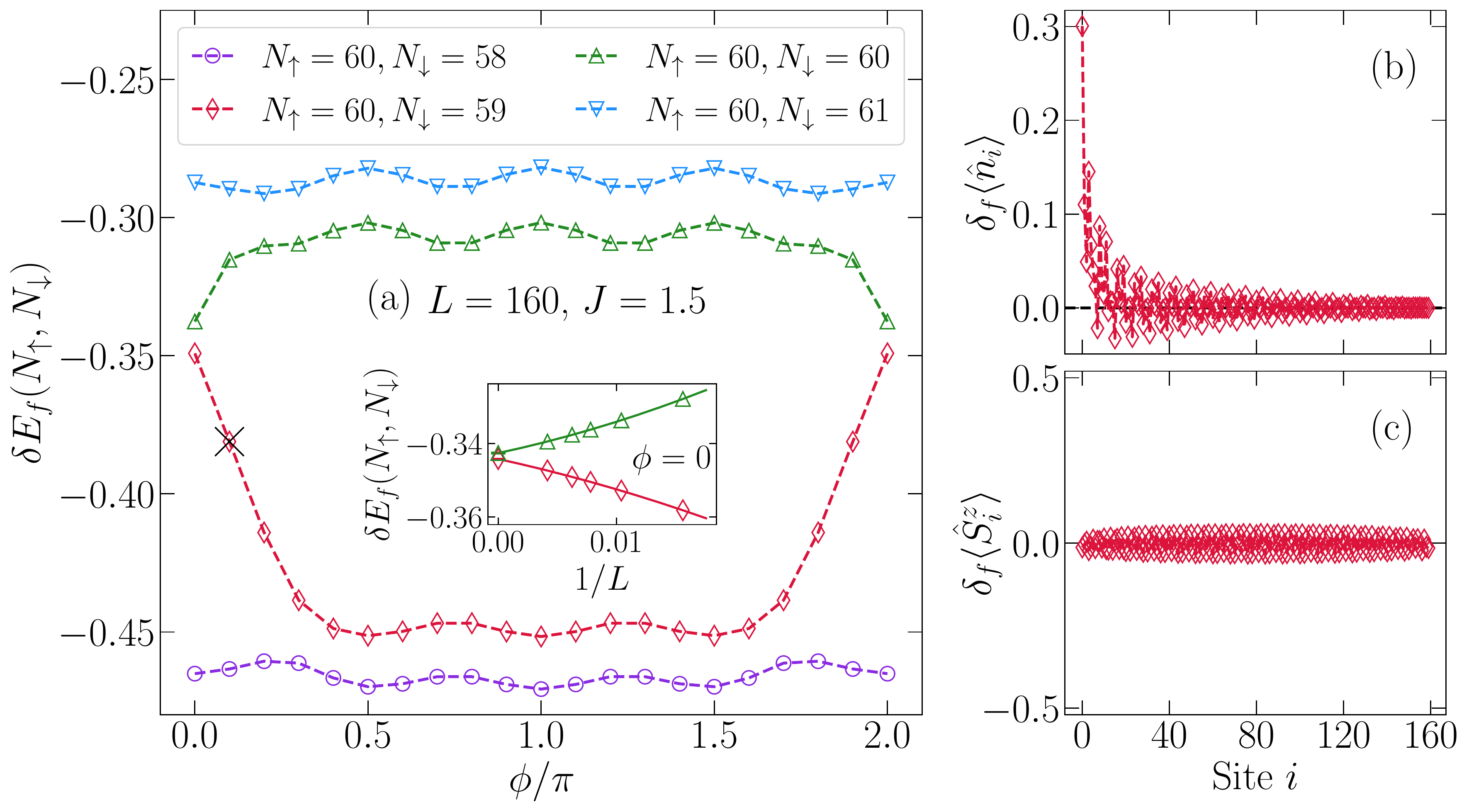}
  \caption{  (a) The quasi-particle (single fermion) spectrum as a function of the phase factor $\phi$ near the filling $\rho=3/4$, the inset displays the finite size extrapolation ($L=64,96,128,160,240$) of $\delta E_f(N_{\uparrow},N_{\downarrow})$ at $\phi=0$. (b) The corresponding onsite charge and spin difference for the two adjacent many-body ground state, at parameters labeled by the black cross in (a). Here results are from calculations of the system with $J=1.5$ and OBCs. All panels share the same legend. }
  \label{fig:topo_3a}
\end{figure}

Since the two insulating regions share similar properties in bulk, we again assume the single fermion as the elementary edge mode for $J>0.85$. Although the quasi-particle spectrum in Fig.~\ref{fig:topo_3a}(a) gives a very different first impression, the overall bulk-edge correspondence is substantially the same as what is shown in Fig.~\ref{fig:topo_1} by shifting the phase factor by $\pi$. However, there is a small remnant gap at $\phi=0$, which is supposed to be the band-touching point. We attribute this bad connection to the finite size effect and display the finite size extrapolation in the inset of Fig.~\ref{fig:topo_3a}(a), in which the remnant gap eventually closes as the system approaches the thermodynamic limit. Therefore, we believe the quasi-particle corresponds to the gapless edge mode is the same in both insulating regions at $\rho=3/4$, and is the same to the case of $\rho=1/4$.

\begin{figure}[!ht] 
  \includegraphics[width=1\columnwidth]{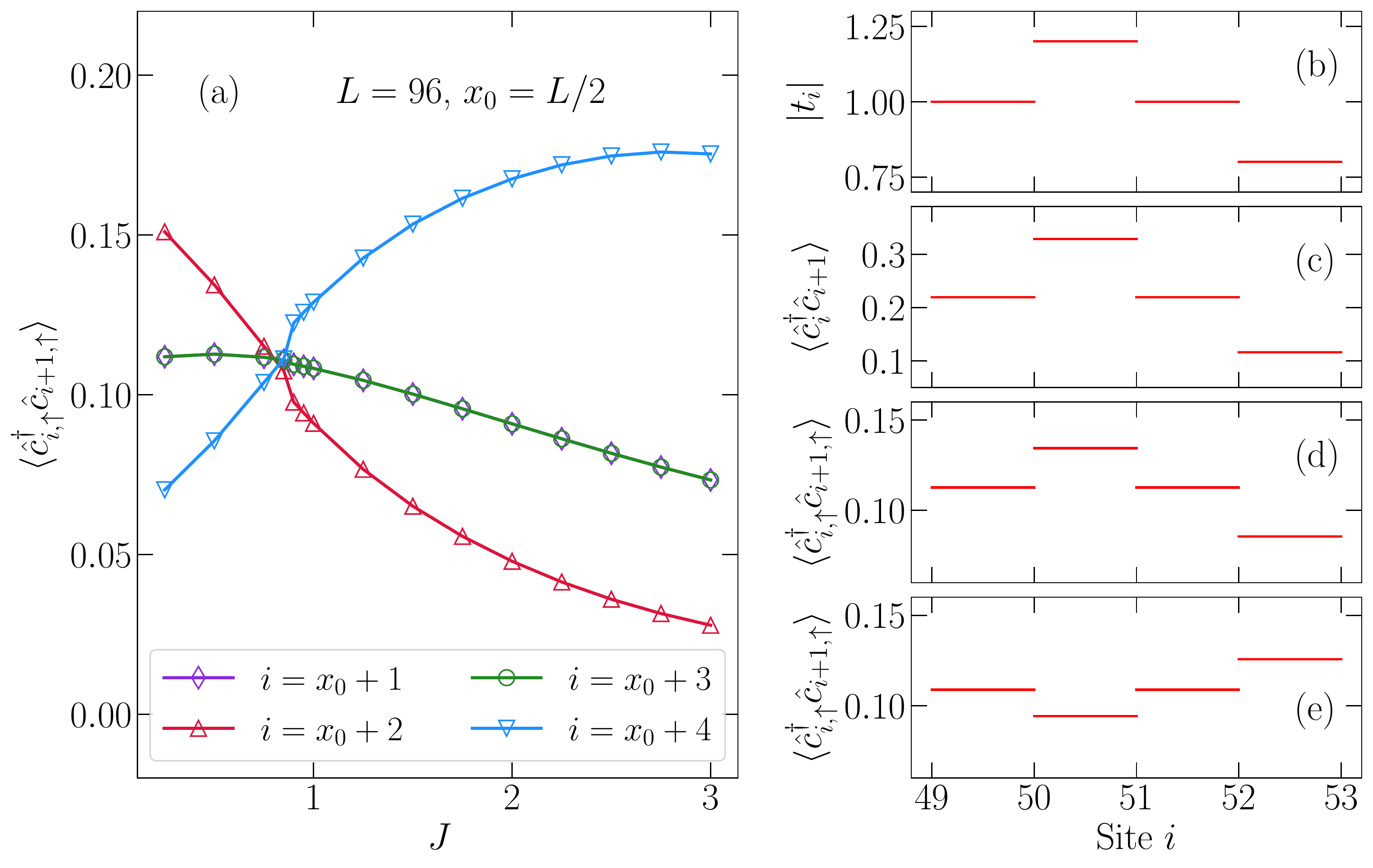}
  \caption{ (a) The effective hopping $\langle \hat{c}^\dagger_{i,\uparrow}\hat{c}_{i+1,\uparrow} \rangle$ at 4 adjacent bonds as a function of exchange interaction $J$. The input hopping (b), the effective hopping at $J=0.5$ (d) and $J=1.5$ (e) on same adjacent bonds. (c) The effective hopping of the noninteracting spinless fermion chain with the same modulation. In all panels we have $\phi=0$, $L=96$ and PBCs. }
  \label{fig:topo_3b}
\end{figure}

There is a remaining issue at this filling that why there is a $\pi$ phase shift in the quasi-particle spectrum across the gap closing point between bulk insulating states, while these two regions have the same Chern number and same edge modes. To answer this question, we turn to the local energy bonds of the ground state, which is demonstrated to be able to detect topological phase transitions in various models possessing symmetry-protected topological phases or long-range topological orders~\cite{Yu2020}. For the quasi-periodic system with period $p=4$, we examine the effective hopping $\langle \hat{c}^\dagger_{i,\uparrow}\hat{c}_{i+1,\uparrow} \rangle$ on 4 different bonds in one unit cell in the middle of the chain. As displayed in Fig.~\ref{fig:topo_3b}(a), effective hopping bonds as functions of $J$ already show signals of a transition at $J\approx0.85$, which agrees with the position the charge gap closes [see Fig.~\ref{fig:energy_diffrho-2}(d)]. We further explicitly compare $\langle \hat{c}^\dagger_{i,\uparrow}\hat{c}_{i+1,\uparrow} \rangle$ with the magnitude of the input hopping $|t_i|$, and which of the noninteracting fermions. For the spinless fermions with the same modulation, in the noninteracting setting, the effective hopping must follow the wave pattern of the input hopping parameters, 
as verified by the data in Fig.~\ref{fig:topo_3b}(b) and (c). In presence of interactions, the profile of $\langle \hat{c}^\dagger_{i,\uparrow}\hat{c}_{i+1,\uparrow} \rangle$ also agrees with the input $t_i$ in most cases, as shown in Fig.~\ref{fig:topo_3b}(d) for $J=0.5$ and $\rho=3/4$ as an example. However, for $J$ above the critical point, the wave pattern shifted by a phase of $\pi$, as shown in Fig.~\ref{fig:topo_3b}(e). This $\pi$-phase shift of the effective hopping accompanied by the bulk-state gap closing directly leads to a $\pi$-phase shift in the quasi-particle spectrum.    

\bigskip
\section{summary and discussion}
\label{sec:summary}

We systematically investigate the ground-state properties of the extended $t$-$J$ model with modulated hopping and interaction. The system has a rich ground-state phase diagram consisting of the metallic state, the superconducting state, the phase separation, and different insulating states with nontrivial topological nature. The superconducting state is characterized by the significant singlet spin gap and negative binding energy, the dominant pairing correlation decay, and the divergent paring structure factor at zero momentum. The spin gap and the magnitude of binding energy have been dramatically enlarged in the superconducting phase, which can be attributed to the flat-band structure of the quasi-periodic superlattice. This enhancement of the superconductivity, which was first proposed in coupled $t$-$J$ segments~\cite{Reja2016}, has been confirmed in a more general quasi-periodic geometry. However, here in this work, the superconducting region in the $\rho$-$J$ plane is not significantly expanded compared to the homogeneous $t$-$J$ chain~\cite{Moreno2011} and is still far from $J\approx 1/3$ in actual materials. 

The interplay among the quasi-periodicity, the interaction, and the 1D confinement leads to various topological edge modes that manifest an extreme spin-charge separation nature. At $\rho=1/4$ and $3/4$, while the spin-1/2 fermion contributes the energy of the gapless edge mode, only the charge degree of freedom manifests the localized excitation at the end. However, the spin excitation completely merges in bulk. At $\rho=1/2$, we demonstrate two bosonic edges modes (related to the magnon and the singlet pair) and the corresponding interaction-driven transition between them. There is also a topological transition at $\rho=3/4$ accompanied by the $\pi$-phase shift in bulk's effective hopping wave pattern.

Our investigation has shown pregnant physics in the 1D $t$-$J$ model in the presence of off-diagonal modulations, with the fixed unit cell size and modulation magnitude. Moreover, there are still plenty of interesting issues to be explored in the modulated spinful fermionic chains. For example, is it possible to find the gapless magnon modes at the edges in the enhanced superconducting state with the significant spin gap? We do not find this new kind of topological superconductivity in the present work with the current setting and parameters, but we cannot rule out the possibility. On the other hand, we expect the rich physics in the modulated fermionic system can stimulate experimental studies in the platform of cold atoms~\cite{Jepsen2020} or the quantum computing devices~\cite{arute2020observation}.

\section*{Acknowledgement}

We thank Z.-S. Zhou for helpful discussions in numerics. This research was supported by the National Natural Science Foundation of China (grant nos. 11904145, 12174167, 11834005, and 12047501). The computations were partially performed in the Tianhe-2JK at the Beijing Computational Science Research Center (CSRC).

\appendix
\section{Correlations between sites \& unit cells}
\label{appendix:A}

\begin{figure}[!t] 
  \includegraphics[width=1\columnwidth]{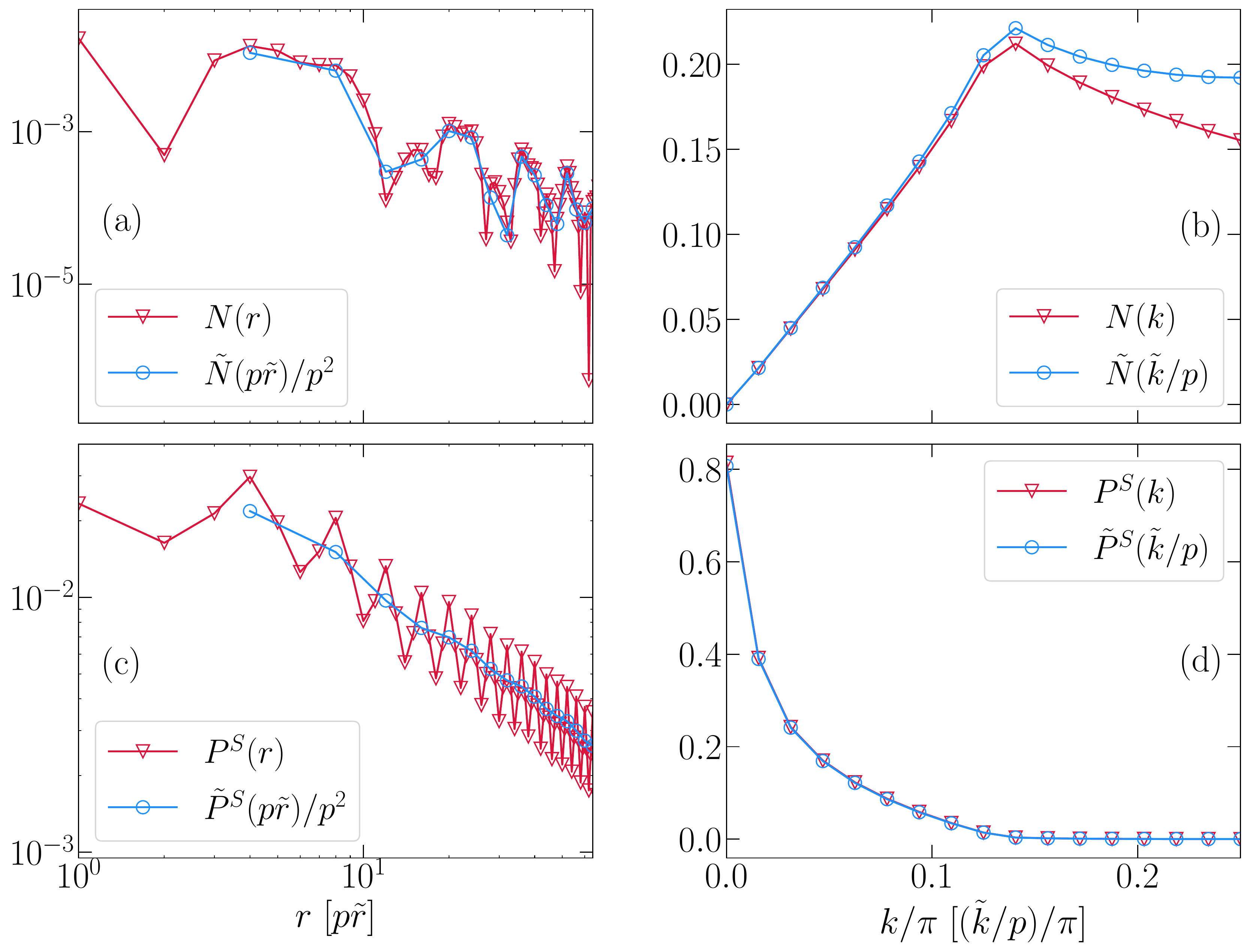}
  \caption{ Direct comparison between correlations based on sites and unit cells. a) and c) The real space decay of the density-density correlation function and the pairing correlation function. b) and d) Corresponding structure factors. Here we use data for $J=2.25$ and $L=128$ at $\rho=1/8$ with OBCs.  }
  \label{fig:appdendix_corr_cells}
\end{figure}

In the main paper we adopt the correlation function $X_{ij}$ between sites $\{i,j\}$. Although it is demonstrated that the correlation function defined between unit cells gives the same Luttinger parameter~\cite{Reja2016,Valencia2002}, in this section, we explicitly compare these two forms of correlations. The correlation function between unit cells can be easily extracted from which between sites as:
\begin{align}
\tilde{X}_{\alpha,\beta} = \sum_{i\in \alpha,j \in \beta} X_{ij},
\end{align}
where $\alpha$ and $\beta$ are indices of unit cells. Similar to Eq.~(\ref{eq:Xr}) and (\ref{eq:Xk}), we define the correlation decay 
\begin{align}
\label{eq:appendix_Xr}
  \tilde{X}(R) = \frac{1}{\cal \tilde{N}}\sum_{|\alpha-\beta|=\tilde{r}}\tilde{X}_{\alpha\beta},
\end{align}
and the structure factor 
\begin{align}
\label{eq:appendix_Xk}
  \tilde{X}(k) = \frac{1}{\tilde{L}}\sum_{\alpha,\beta=1}^{\tilde{L}}\mathrm{e}^{\mathrm{i}\tilde{k}(\alpha-\beta)} \tilde{X}_{\alpha\beta}.
\end{align}
Here $\tilde{L}=L/p$ is the number of unit cells, and $\tilde{r}$ ($\tilde{k}$) is the distance in the real (momentum) space based on the unit cell. By the straightforward mapping $\tilde{r}p \rightarrow r$ and $\tilde{k}/p \rightarrow k$, we can directly compare the correlation and the corresponding structure factor based on sites and unit cells. In Fig.~\ref{fig:appdendix_corr_cells}, we present typical numerical results for the density-density and pairing correlation function, as well as corresponding structure factors. After being rescaled, curves extracted from correlations between sites and unit cells show very close behaviors. Differences in local details can be attributed to the fact that the correlation based on unit cells neglects 
terms within the same unit cell.  


\section{Phase separation}
\label{appendix:B}

\begin{figure}[!b] 
  \includegraphics[width=1\columnwidth]{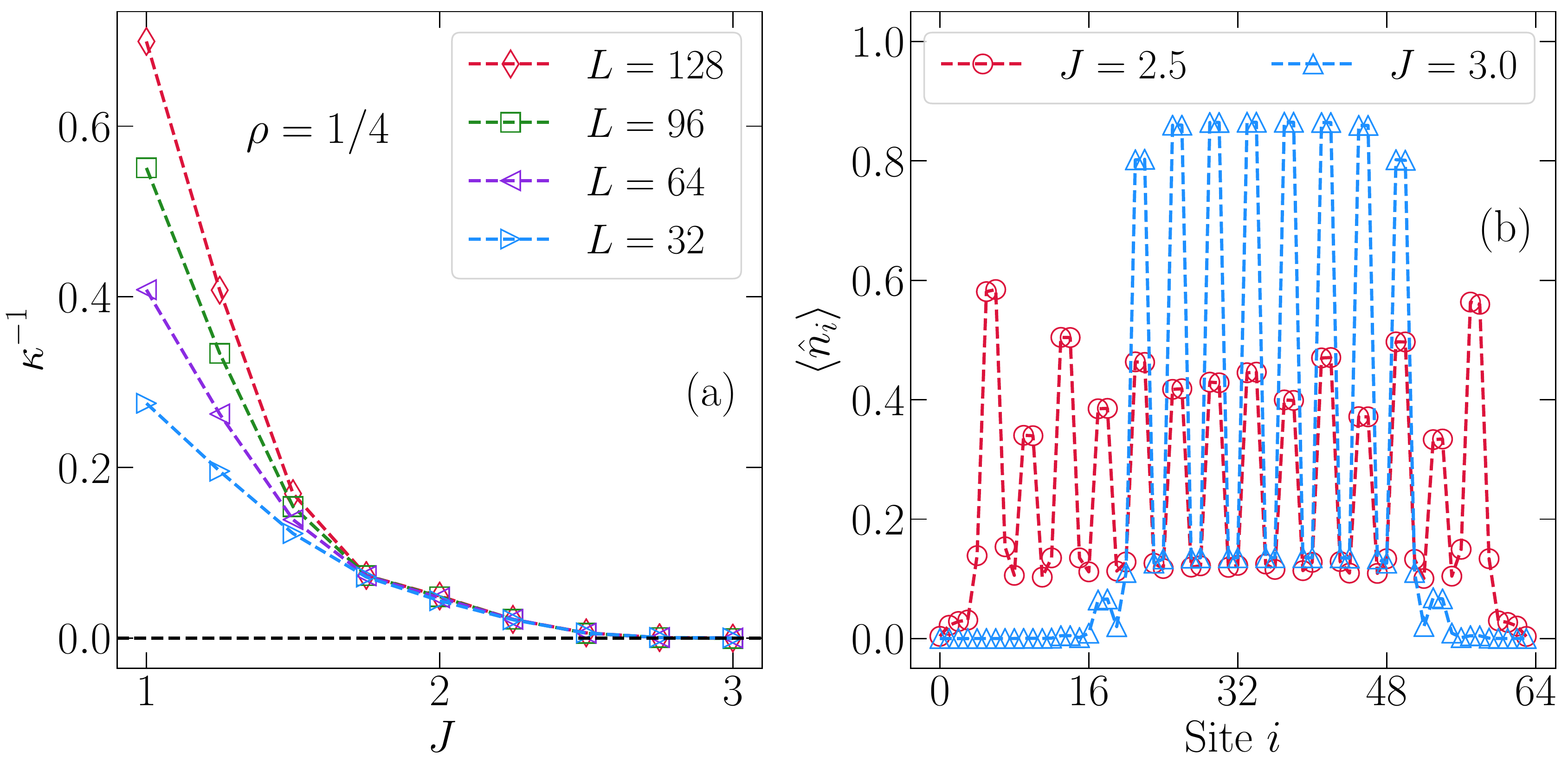}
  \caption{  a) The inverse compressibility $\kappa^{-1}$ as a function of $J$ at $\rho=1/4$ for different system size $L$. b) The charge distribution in the conducting state ($J=2.5$) and the phase separation ($J=3.0$) with $L=64$. Here OBCs are adopted.  }
  \label{fig:appdendix_PS}
\end{figure}

For the $t$-$J$ model, large exchange interaction $J$ leads to the unstable phase separation~\cite{Moreno2011}, which can be detected by the inverse compressibility
\begin{align}
  \kappa^{-1}(\rho) &= \rho^2 \frac{\partial^2 e_0(\rho)}{\partial \rho^2} \nonumber \\
  &\approx \rho^2 \frac{e_0(\rho + \Delta \rho) + e_0(\rho - \Delta \rho) - 2 e_0(\rho)}{\Delta \rho^2},
\end{align}
where $e_0(\rho) = E_0/L$ is the ground-state energy per site. In this work, we take not extra calculations for the phase separation state, since which is not our main focus. Instead, we adopt a special $\Delta\rho=2/L$ and therefore can extract $\kappa^{-1}$ from the charge gap $\Delta E_C$ as 
\begin{align}
    \label{eq:appdendix_B2}
    \kappa^{-1}(N/L) \approx \frac{N^2}{2L}\Delta E_C(L,N).
\end{align}
The estimated $\kappa^{-1}$ from Eq.~(\ref{eq:appdendix_B2}) is displayed in Fig.~\ref{fig:appdendix_PS}(a), for $\rho=1/4$ as an example. For small $J$s, The inverse compressibility increases as the system size increases, which ensures the system is away from phase separation with positive $\kappa^{-1}$. Roughly at $J\geq2.75$, the inverse compressibility shows less system size dependency and approaches $0$. This estimated critical point can be verified by the density profile shown in Fig.~\ref{fig:appdendix_PS}(b). For an exchange interaction slightly larger than the critical point, the system separates into particle-rich and hole-rich regions in real space, and the density distribution loses mirror symmetry because of massive degeneracy. By taking the estimation in Eq.~(\ref{eq:appdendix_B2}) at different densities, we obtain the phase boundary in the ground-state phase diagram~\ref{fig:phase_diagram} in the main paper.

\bibliography{modulated_tJ_refs}

\end{document}